\documentclass[fleqn,twoside]{article}
\usepackage{espcrc2}
\usepackage{graphicx}

\newcommand{\AmS}{{\protect\the\textfont2
  A\kern-.1667em\lower.5ex\hbox{M}\kern-.125emS}}

\title{Vortex configurations and metastability in
mesoscopic superconductors}

\author{Cl\'ecio C. de Souza Silva\thanks{Present Address: Laboratory for Solid State
Physics and Magnetism, Katholieke Universiteit Leuven,
Celestijnenlaan 200D, B-3001 Leuven, Belgium}, Leonardo R.E.
Cabral\thanks{Present Address: Departement Natuurkunde,
Universiteit Antwerpen (UIA), B-2610 Antwerpen,
Belgium}, J. Albino Aguiar\\
\addressmark{Departamento de F\'{\i}sica, Universidade
Federal de Pernambuco, 50670-901 Recife, PE, Brasil}}

\begin{document}



\begin{abstract}

The vortex dynamics in mesoscopic superconducting cylinders with
rectangular cross section under an axially applied magnetic field
is investigated in the multivortex London regime. The rectangles
considered range from a square up to an infinite slab. The flux
distribution and total flux carried by a vortex placed in an
arbitrary position of the sample is calculated analytically by
assuming Clem's solution for the vortex core. The Bean-Livingston
energy barrier is also analytically calculated in this framework.
A Langevin algorithm simulates the flux penetration and dynamical
evolution of the vortices as the external field is slowly cycled.
The simulated magnetization process is governed by metastable
states. The magnetization curves are hysteretic, with paramagnetic
response in part of the downward branch, and present a series of
peaks corresponding to the entry or expulsion of a single vortex.
For elongated rectangles, the vortices arrange themselves into
parallel vortex chains and an additional modulation of the
magnetization, corresponding to creation or destruction of a
vortex chain, comes out.

\vspace{1pc} Key words: Vortex patterns, surface barrier,
mesoscopic, vortex dynamics \vspace{1pc}

\end{abstract}

\maketitle

\section{Introduction}

Since the pioneering work of Abrikosov~\cite{abrikosov}, it is
well known that in a macroscopic type-II superconductor the
magnetic field, in the so-called mixed state, penetrates in the
form of singly quantized flux lines (or vortices), which form a
triangular lattice. For finite samples the vortices are formed at
the surface and then pulled to the sample center by the shielding
Meissner currents. Notwithstanding, close to the sample edge, a
vortex is strongly attracted by the superconductor-–vacuum
interface. These competing interactions give rise to a energy
barrier, the so-called Bean-Livingston surface barrier, which
retards the movement of vortices towards the sample
center~\cite{BL,degennes}. In samples with very smooth interfaces,
this barrier leads to irreversible field-dependent magnetization
loops $M(H)$ and finite critical currents even in the absence of
inhomogeneities~\cite{ClemSB,burlach}. Nevertheless, defects in
the interface may cause the local destruction of the surface
barrier, opening leaks for vortex entry.

Recent progress in nanostructuring of superconducting materials
\cite{geim,victor} has provided the opportunity of studying the
vortex state in mesoscopic samples, i.e. whose dimensions are of
the order of the penetration depth  $\lambda$ and/or the coherence
length $\xi$, with sharp interfaces and no noticeable pinning. At
these length scales the vortex lattice and the vortex itself may
present new and very interesting properties, as for example, the
formation of multi-quanta giant vortices, vortex molecules and
chain-like vortex arrangements. These vortex structures are
strongly dependent on the sample geometry and size, as has been
shown by numerical and analytical solutions of the Ginzburg-Landau
(GL) equations~\cite{peeters,victor00} and
experiments~\cite{geim,victor00}, and may appear even in samples
made of type-I materials (with $\kappa=\lambda/\xi<1/\sqrt{2}$),
such as aluminium. The reason is that in thin films the parameter
governing the appearance of vortices is actually the effective GL
parameter, $\tilde{\kappa}=\Lambda/\xi$, where
$\Lambda=\lambda^2/d$ may be much greater than $\lambda$ for small
sample thicknesses $d$. The choice between the multivortex and the
giant vortex states depends on the value of
$\tilde{\kappa}$~\cite{peeters98} and on the system size as well.
Here we shall consider only systems with high $\kappa$ and sizes
comparable to $\lambda$, though larger than $\xi$. For these
systems the multivortex state persists over a wide area of the
phase diagram.

In the course of our recent research with mesoscopic
superconductors~\cite{cl01,cl02,cl03} we have been studying the
vortex dynamics, within the London approximation, in films,
multilayers, and strips, submitted to an external magnetic field
and dc currents. In this article we present new results for
mesoscopic rectangles in the presence of an applied magnetic
field. The vortex structure and the energy surface barrier present
in these systems are described within the London limit of the
Ginzburg-Landau equations. The magnetization process is simulated
by a fast algorithm based on Langevin dynamics. We observe, for
elongated rectangles, the appearance of vortex chain states and
transitions between these states resulting in a modulation of the
magnetization loop. Particular attention is given to the formation
of metastable vortex configurations in these systems during the
magnetization process, which leads to hysteretic magnetization
curves with paramagnetic response in part of the downward branch.

\section{Vortex structure and the Bean-Livingston barrier}

In the high-$\kappa$ limit, the superconducting order parameter is
essentially homogeneous except near a vortex core, where its
spatial distribution may be given approximately by~\cite{core}
\begin{equation} \label{eq:clem}
    \psi(\rho) \approx \frac{\rho e^{i\varphi}}
    {(\rho^2 +2\xi^2)^{1/2}}\,,
\end{equation}
where $\rho$ is the distance to the vortex center. For points
${\bf r} = (x,y)$ far away from the vortex core, located at ${\bf
r}' = (x',y')$, $|\psi({\bf r})|$ is uniform and the second
Ginzburg-Landau (GL) equation reduces to the London equation ,
\begin{equation} \label{eq:London}
    -\lambda^2\nabla^2b(x,y) + b(x,y) =
    \phi_0\delta({\bf r-r'})\,.
\end{equation}
The solution near the vortex core may be accomplished very
precisely by inserting the variational trial function
(\ref{eq:clem}) into the GL free energy functional. The result is
equivalent to making the cutoff $b(r)\rightarrow
b(\sqrt{r^2+2\xi^2})$ in the London solution for the vortex local
induction.

In this spirit, one may compute the flux distribution of a vortex
confined in a cylinder if the appropriate boundary condition is
used. Here we consider long cylinders with a rectangular cross
section of width $W$ and length $L$. An external magnetic field
$H$ is applied axially and the vortices are assumed to be
perfectly aligned with ${\bf H}$. This problem has been considered
previously within the London theory by Sardella, Doria and
Netto~\cite{sardella}. Here we shall use Clem's variational
solution for the vortex core to compute the local field
distribution and the position-dependent, effective flux carried by
a vortex. The result for the local induction generated by a vortex
is
\begin{eqnarray}
    b_{v}(x,y) &=& \frac{\phi_0}{\lambda W}\sum^{\infty}_{m=1}
    \frac{\cos k_mx_- - \cos k_mx_+}{k_m}
    \times \nonumber \\
& & \frac{\cosh k_m(L-y_-) - \cosh k_m(L-y_+)}{\sin k_mL}
    \nonumber \\
\end {eqnarray}
where $k_m = m\pi/W$, $\alpha_+ = \alpha + \alpha'$ and $\alpha_-
= |\alpha - \alpha'| + \xi$ ($\alpha = x,y$). With this solution,
one can compute the interaction energy between vortex $i$ and $j$
in the usual way, that is, $E_{vv}(x_i,y_i;x_j,y_j) =
\phi_0b_v^{(j)}(x_i,y_i)$. Integrating the local induction
$b_v(x,y)$ over the entire sample area, one finds the effective
magnetic flux carried by a vortex,
\begin{eqnarray}
    \phi(x,y) &=& \phi_0\Big[1-\frac{\cosh(x/\lambda - W/2\lambda)}
    {\cosh W/2\lambda} - \frac{4}{\lambda^2W} \times\nonumber \\
& & {\sum_{m=1}^\infty}'
    \frac{\sin k_mx}{k_m^3}\frac{\cosh k_m(y-\frac{L}{2})}
    {\cosh k_mL/2}\Big]\,.
\end{eqnarray}
The prime in the sum operator indicates that only the terms where
$m$ is odd are taken into account. Note that in mesoscopic samples
the effective flux may be quite smaller than the flux quantum,
even for a vortex positioned at the sample center, where
$\phi(x,y)$ is maximum.

The homogeneous solution of Eq.~\ref{eq:London} corresponds to the
local Meissner screening flux distribution, which is given by
\begin{eqnarray}
    b_M(x,y) &=& H\Big[\frac{\cosh(x/\lambda - W/2\lambda)}
    {\cosh W/2\lambda} + \frac{4}{\lambda^2W} \times\nonumber \\
& & {\sum_{m=1}^\infty}'
    \frac{\sin k_mx}{k_m^3}\frac{\cosh k_m(y-\frac{L}{2})}
    {\cosh k_mL/2}\Big] \nonumber \\
&=& H\Big[1-\frac{\phi(x,y)}{\phi_0}\Big]\,.
\end{eqnarray}
This field distribution is minimum at the samples center. Thus,
the Meissner screening current density, ${\bf j}_M = \mu_0^{-1}
\nabla\times{\bf b}_M$, repels the vortex away from the surfaces,
towards the sample center. The corresponding potential energy felt
by the vortex is then given by $E_M = \phi_0b_M(x',y')$.

On the other hand, there is an energy cost to put a vortex inside
the superconducting specimen. This is given by the vortex
self-energy
\begin{equation} \label{eq:Eself}
    E_{self} = \frac{1}{2}E_{vv}({\bf r=r}').
\end{equation}
This energy depends on the vortex position and is maximum at the
sample center. That is, the vortex is attracted by the surfaces
or, in other words, by its images, which are necessary to satisfy
the boundary conditions. The attractive self-energy and the
repulsive Meissner energy form the well known surface
(Bean-Livingston) barrier, which delays vortex entrance and exit
in the superconducting sample.

\begin{figure}[htb]
\begin{center}\leavevmode
\includegraphics[width=0.85\linewidth]{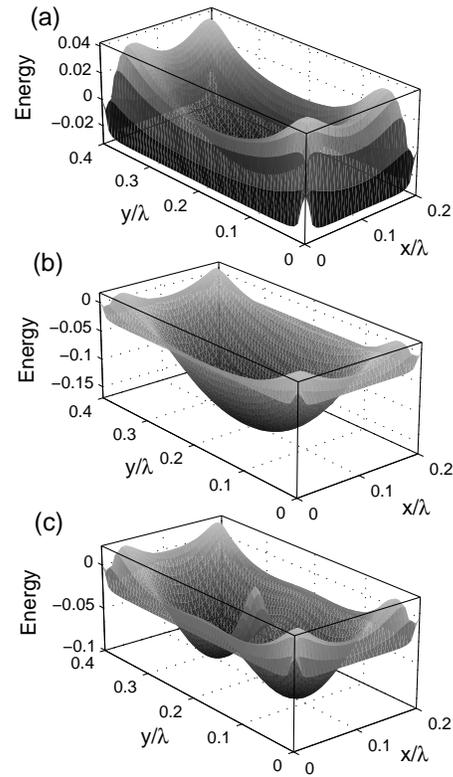}
\vspace{-5mm} \caption{(a) and (b): Energy distribution in a
superconducting rectangle with $W=0.2\lambda=20\xi$ and $L=2W$ in
the absence of vortices at the first critical field, $H=H_{c1}$
(a), and at the entrance field, $H=H_{en}$ (b). (c) The total
energy when a vortex is located at the sample center at
$H=H_{en}$.} \label{fig1}
\end{center}
\end{figure}

In Fig.~\ref{fig1}, we make a surface plot of the energy
distribution in a superconducting rectangle with
$W=0.2\lambda=20\xi$ and $L=2W$ for three different situations. In
Fig.~\ref{fig1}(a), H equals the lower critical field, $H=H_{c1}$,
for which there is a global minimum at the sample center and a
strong energy barrier near the sample surfaces. In mesoscopic
samples the activation energy to overcome this barrier is usually
much higher than the thermal energy $k_BT$. In Fig.~\ref{fig1}(b),
$H=H_{en}$, where the surface barrier for the first vortex entry
disappears. The entrance field may be defined as
$\partial{E}_{self}/\partial{x} =
-\partial{E}_M(H_{en})/\partial{x}$, with the derivatives being
evaluated at one of the points $(x,y)=(\xi,L/2)$ or $(D-\xi,L/2)$,
as suggested by Fig.~\ref{fig1}(b). Note that in the smaller edges
of the sample, the barrier is still present, which means that in a
rectangle the first vortex tends to enter the sample through one
of the larger edges. The total energy when a vortex is placed at
the sample center at $H=H_{en}$ is depicted in Fig.~\ref{fig1}(c).
In this case, a new energy barrier for the entrance of a second
vortex is developed. This means that, at zero temperature, the
second vortex will be allowed to enter the sample only at a field
$H_2>H_{en}$. In the field region $H_{en}\leq H<H_2$, the sample
behaves as in the Meissner state.

\section{Langevin dynamics and simulation scheme}

Next, we review our model to study vortex dynamics in confined
geometries. The time evolution of a vortex is described by
overdamped Langevin equations of motion,
\begin{equation} \label{eq:Eself}
    \eta{\bf v}_i = -\nabla_i E +
    \mbox{\boldmath$\Gamma$}_i(t)\,,
\end{equation}
where
\begin{eqnarray} \label{eq:Etot}
    E &=& \frac{1}{2}\sum_{i,j}E_{vv}({\bf r}_i,{\bf r}_j)+
    \nonumber\\
& & \sum_i\big[E_{self}({\bf r}_i) + E_M({\bf r}_i)\big]
\end{eqnarray}
is the total energy of the vortex distribution, $\eta$ is the
Bardeen-Stephen friction coefficient, ${\bf v}_i$ is the vortex
$i$ velocity. $\mbox{\boldmath$\Gamma$}_i$ is a Gaussian
stochastic noise related with a small temperature $T$ and the
friction $\eta$ by
$\langle\Gamma_{\alpha,i}(t)\Gamma_{\beta,j}(t')\rangle = 2\eta
k_BT\delta_{\alpha\beta}\delta_{ij}\delta(t-t')$, where the Greek
indices stand for the directions $x$ or $y$. This temperature
plays the role of accelerating the convergence towards a
stationary state.

The simulation consists in numerically integrating the coupled
Langevin equations for all vortices inside the sample. The
integration is based on a finite difference algorithm. A vortex is
allowed to enter the sample and participate on the dynamics if it
satisfies a force balance condition near one of the sample
surfaces. The procedure is the same adopted in earlier
publications \cite{cl01,cl02,cl03}: at each time step a test
vortex is placed in a random position a distance $\xi$ from one of
the sample interfaces and the total force acting on it is
calculated. If this force points to the sample interior, the
vortex is accepted, otherwise it is rejected.

The magnetization loops are calculated as an external field is
varied in small steps during up to $5\times 10^5$ time steps. We
verified that the time between successive field changes is large
enough for the system to relax to the desired stationary states.
The magnetization $M=\mu_0^{-1}B-H$ is calculated by integrating
the total amount of magnetic induction $B$ inside the sample,
$B=\frac{1}{WL}\big[ \int\!dx\,dy\,b_M(x,y) +
\sum_i\phi(x_i,y_i)\big]$.

\section{Vortex chain states in mesoscopic films}

\begin{figure}[thbp]
\begin{center}\leavevmode
\includegraphics[width=0.98\linewidth]{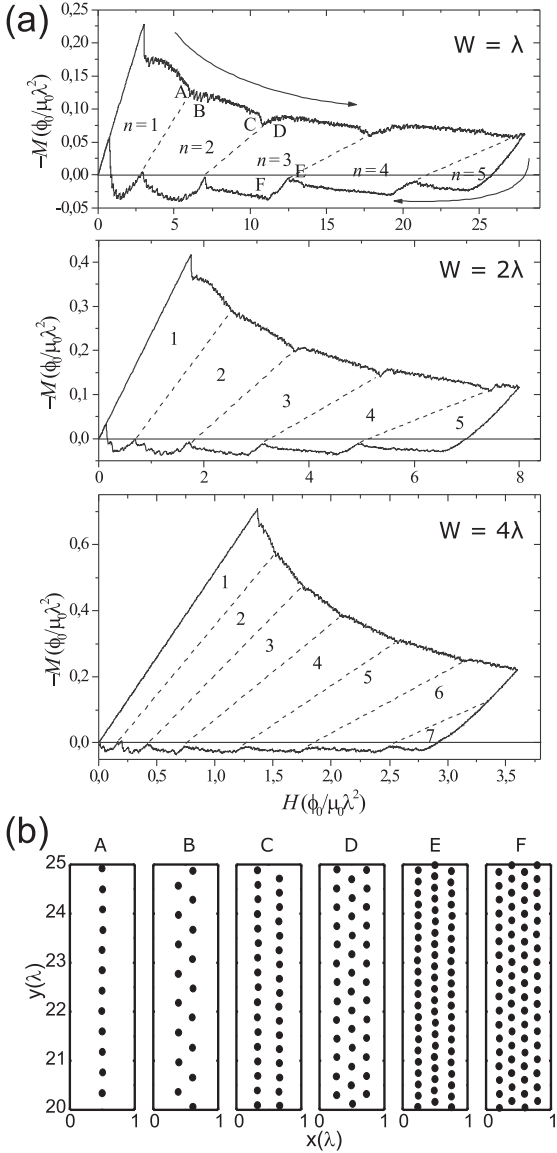}
\vspace{-5mm}\caption{ (a) Field dependent magnetization loops for
infinite films of width $W=\lambda$, $2\lambda$ and $4\lambda$.
The dashed lines are just guides for a better identification of
the transitions between successive vortex chain states. (b)
Snapshots of the vortex lattice in the points of the magnetization
loop indicated at the $W=\lambda$ film magnetization curve.}
\label{fig2}
\end{center}
\end{figure}

Here, we consider the case where $L\rightarrow\infty$, which
corresponds to a film infinitely long in the $y$ and $z$
directions. To perform the simulation, we take a slice $L_y$ in
the $y$ direction and assume periodic boundary conditions at
$y=0,L_y$, in such a way that the simulation is restricted to the
region $0\leq x\leq W$, $0\leq y\leq L_y$.

In Fig. \ref{fig2}(a) we show magnetization loops for several film
thicknesses, $W/\lambda$ = 1, 2 and 4. We choose $\kappa=20$ and
$L_y = 40\lambda$. All the loops are characterized by hysteretic
behavior and a distribution of peaks at specific field values.
Snapshots of the vortex configurations [some of which are depicted
in Fig.~\ref{fig2}(b)] show that the VL is composed of $n$ linear
chains of vortices parallel to the film surfaces and the extra
magnetization peaks are associated with sudden rearrangements of
the VL to accommodate a new vortex chain ($n\rightarrow n+1$
transitions), in the case of increasing magnetic field, or destroy
an existent chain ($n\rightarrow n-1$ transitions), for decreasing
field. Note that as the film thickness increases the matching
peaks become less evident in the magnetization loops, that is, the
film gradually crosses over from mesoscopic to macroscopic
behavior. The matching effect of vortex chains in thin slabs
($W<\lambda$) has been studied in the framework of thermodynamic
equilibrium\cite{mefilm}. Our approach assumes that a vortex
nucleates at the surfaces, i.e., for fields just above $H_{c1}$ it
has to overcome a strong energy barrier in order to find the
global minimum at the equatorial plane of the film. Hence we are
dealing with steady metastable states. This situation is closer to
real experimental conditions, where the time necessary for the
system to relax to the thermodynamic equilibrium is not
accessible. In this case the critical fields where the transitions
$n\rightarrow n\pm 1$ take place are history dependent. In Fig.
\ref{fig2}(b) we show snapshots of the vortex configuration in a
slice of the $W=\lambda$ film for different points of the
magnetization cycle indicated in the upper panel of Fig.
\ref{fig2}(a).

\section{Vortex states in mesoscopic rectangles}

\begin{figure}[thb]
\begin{center}\leavevmode
\includegraphics[width=.97\linewidth]{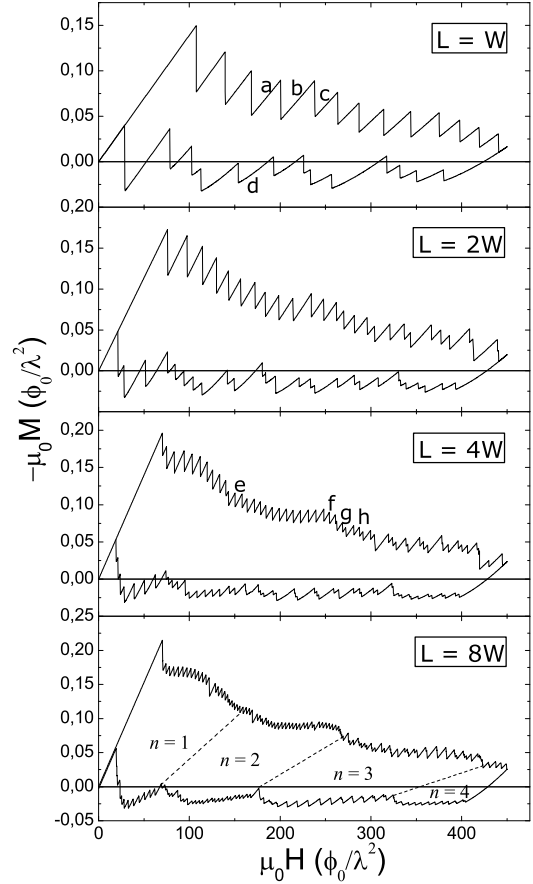}
\vspace{-7mm} \caption{ (a) Magnetization loops of rectangles of
different aspect ratio, $W/L$ = 1/1, 1/2, 1/4, and 1/8, with
$W=0.2\lambda =20\xi$.} \label{fig3}
\end{center}
\end{figure}

\begin{figure}[thb]
\begin{center}\leavevmode
\includegraphics[width=\linewidth]{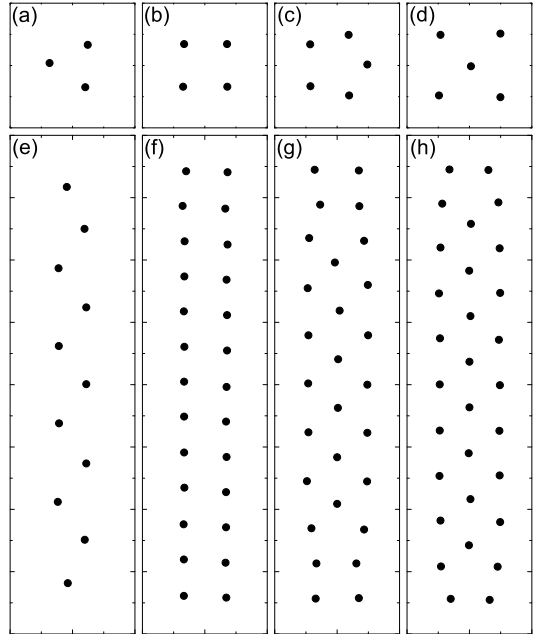}
\vspace{-5mm} \caption{Vortex configurations in a square (a-d) and
for a rectangle with $L=8W$ (e-g), both considering
$W=0.2\lambda=20\xi$. The points in the magnetization curve
corresponding to these configurations are indicated in
Fig.~\ref{fig3}} \label{fig4}
\end{center}
\end{figure}

Now we consider the more general case of long cylinders with
rectangular cross section. The calculations were performed for
different aspect ratios, $W/L$ = 1/1, 1/2, 1/4, and 1/8, with $W$
fixed at $W=0.2\lambda$ and $\kappa = 100$. The magnetization
loops for these samples are depicted in Fig.~\ref{fig3} and in
Fig.~\ref{fig4} we show some vortex configurations for the $L=W$
and $L=4W$ samples. The magnetization peaks correspond to entry
(in the upward magnetization branch) or expulsion (in the downward
branch) of an individual vortex. In some points of the
magnetization curve of the longer samples ($L=4W$ and $L=8W$),
jumps of two or more vortices are also observed, specially at
higher fields.

States with same vorticity may have two distinct metastable
solutions, depending on whether the field is going up or down.
This is illustrated in Figs.~\ref{fig4}(c) and (d), where in the
state with vorticity five in a square, vortices arrange in a
pentagon, in upward field, or in a face centered square, in
downward field.

As the rectangle is elongated the vortices organize into parallel
chains of vortices resembling the chain states in infinite films.
The effect of these arrangements are seen as a modulation of the
magnetization loop. Note that the magnetization curve of the
$L=8W$ rectangle is particularly similar to that of the
magnetization of the $W=\lambda$ infinite film. In both systems
the ratio $W/\xi = 20$.

As illustrated in Figs.~\ref{fig2} and \ref{fig3}, hysteresis in
the magnetization curve is a property shared by all the samples
studied here and is a result of the different circumstances in
which vortices enter or are expelled from the sample through the
surface barrier. The surface barrier delays the incursion of
vortices towards the sample interior. As a result, in increasing
magnetic field, the vortex lattice is driven into successive {\it
superheated} metastable states and the diamagnetic response is
stronger than the expected for the thermodynamic equilibrium. In
decreasing field, a surface barrier is also present, now against
vortex expulsion. In this case the system is driven into {\it
supercooled} metastable states and the response is paramagnetic in
most part of the downward branch of the magnetization loop. It is
interesting to note that this behavior of the magnetic response,
diamagnetic in upward field and paramagnetic in part of the
downward branch, is quite frequent in superconducting mesoscopic
samples of different geometries and in a broad range of $\kappa$
values~\cite{geim98,peeters98,victor97}. It has been shown that
this may be explained in terms of flux capture due to the sample
boundaries~\cite{victor97}. Here, Figs.~\ref{fig2} and \ref{fig3}
suggest that this effect may persist even for macroscopic
homogeneous samples.

\section{Conclusions}

In conclusion, we have presented a theoretical study of the
surface barrier in mesoscopic superconducting samples and its
effect on the formation of metastable vortex structures. Using a
Langevin dynamics simulation, magnetization loops of mesoscopic
rectangles of several aspect ratios were calculated. All these
loops are hysteretic and present series of magnetization jumps due
to penetration or expulsion of a vortex. An additional modulation
of the magnetization curve is caused by the formation of vortex
chain states in the more elongated rectangles. Due to the surface
barrier, the vortex configurations are metastable in both branches
of the magnetization loop. The capture of vortices in the downward
branch leads to paramagnetic response, even for the largest
samples studied.

\section*{Acknowledgements}
 This work was supported by the Brazilian Agencies Facepe, CAPES and
 CNPq.

%
%

\end{document}